\documentclass[12pt,aps,epsf]{article}

\usepackage{amssymb,epsfig}

\newcommand{\beq}[1]{
\begin{equation}\label{#1}}
\newcommand{\eeq}{\end{equation}}
\newcommand{\bea}[1]{
\begin{eqnarray}\label{#1}}
\newcommand{\eea}{\end{eqnarray}}
\usepackage{psfrag}

\psfrag{p}{$p$}  
\psfrag{pp}{$p'$}
\psfrag{q}{$q$}
\psfrag{qp}{$q'$}
\psfrag{pip}{$\pi^+$}
\psfrag{pim}{$\pi^-$}
\psfrag{pr}{$P$}
\psfrag{apr}{$\bar P$}
\psfrag{g}{$\gamma$}
\psfrag{gs}{$\gamma^*$}
\psfrag{u}{$u$}
\psfrag{db}{$\bar d$}
\psfrag{ep}{$e^+$}
\psfrag{em}{$e^-$}
\psfrag{d}{$d$} 
\psfrag{a}{$a$}
\psfrag{b}{$b$}
\psfrag{pi}{$\pi$}
\psfrag{k}{$k$}
\psfrag{DA}{$DA$}
\psfrag{TDA}{$TDA$}
\psfrag{TH}{$T_H$}
\newcommand{\out}{\raise-3pt\hbox{\scriptsize    out}}

\begin{document}

\begin{flushright}
   CPHT--RR 023.0405 \\
   hep-ph/0504255
\end{flushright}

\begin{center}  

\textbf{\LARGE QCD analysis of $\bar p N \to \gamma ^* \pi $
in the scaling limit. 
           }

\vspace{0.3cm}

{\large  B. Pire$^1$ and  L. Szymanowski$^{2,3}$  } 

\vspace{0.3cm}

{$^1$CPhT,
\'Ecole  Polytechnique,  F-91128  Palaiseau,  France  \\  
$^2$  Soltan
Institute  for   Nuclear  Studies,  Warsaw,   Poland  \\  
$^3$ Universit\'e  de Li\`ege,  B4000  Li\`ege,
Belgium}

%{\it DRAFT, April 15th, 2005}

\vspace{0.5cm}
\textbf{Abstract}\\
\vspace{1\baselineskip}
\parbox{0.9\textwidth}{
We   study   the   scaling  regime   of  nucleon   -
anti-nucleon  annihilation into a  deeply virtual  photon and  a meson,  
$\bar p N \to \gamma ^* \pi $, in the forward kinematics, where
$|t|<< Q^2 \sim s$.
 We obtain the leading twist amplitude
in the kinematical region where  it factorizes into 
an antiproton distribution amplitude,
a
short-distance matrix element related to 
nucleon form factor and the long-distance dominated 
transition distribution amplitudes  which describe the nucleon to
meson transition. 
We give the $Q^2$  evolution equation for these transition 
distribution amplitudes. 
%We get the
%chiral limit of these distribution amplitudes 
%when the emitted pion is soft. 
The  scaling
of the cross section of this process may be
tested at the proposed GSI intense anti-proton beam facility FAIR with
the PANDA or PAX detectors.
We comment on related processes such as $\pi N \to N' \gamma ^*$
and $\gamma ^* N \to N' \pi$ which may be experimentally studied 
at intense meson beams
 facilities and at JLab or Hermes, respectively.}
\end{center}

 \section{Introduction}

In a recent paper \cite{PS}, we have shown that  factorization 
theorems \cite{fact} 
for exclusive processes apply to the case of the reaction
$\pi^-\,\pi^+ \, \to \, \gamma^*\,\gamma$
in the kinematical regime where the virtual photon is highly virtual (of 
the order of the energy squared of the reaction) but the momentum transfer $t$
is small. We also advocated the extension of this approach to the reaction
$\bar p p \to \gamma ^* \gamma $ and to virtual Compton scattering in 
backwards kinematics. This enlarges the successful description of deep exclusive
reactions  in  terms   of  distribution  amplitudes (DA) \cite{ERBL}  and/or
generalized  parton distributions (GPD) \cite{Dvcs, Dvcs2}  
on the  one  side and
perturbatively   calculable  coefficient  functions   describing  hard
scattering at  the partonic  level on the  other side.  
We want here to describe 
along the same lines the reaction
\beq{pbarp}
%    \begin{equation} 
\bar p(k) N(p) \to \gamma ^*(q) \pi(p') 
%  \label{pbarp}
%\end{equation}
\eeq 
in the  near forward region and for large virtual photon invariant mass 
$Q$, which may  be studied in
detail  at GSI \cite{Panda}.
Such an extension of the GPD framework has already
been advocated in the pioneering work of \cite{FPPS}.
\begin{center} 
% \vspace{1cm}
 % \vspace{1cm}
%%%%%%%%%%%%%%%%%%     FIGURE 1          %%%%%%%%%%%%%%%%%%%%%%%%%%%%
\begin{figure}[t]
\centerline{\epsfxsize7.0cm\epsffile{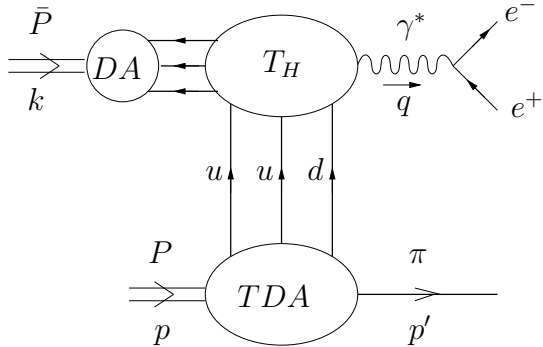}}
\caption[]
{\small The factorization of the annihilation process $\bar p \;p\to \gamma^*
\,\pi$ into the antiproton distribution amplitude $(DA)$, 
the hard subprocess amplitude $(T_H)$ and a baryon $\to$ meson transition
distribution amplitude $(TDA)$ .
 }
%\label{fig:1}
\end{figure}
\end{center}
%%%%%%%%%%%%%%%%%%%%%%%%%%%%%%%%%%%%%%%%%%%%%%%%%%%%%%%%%%%%%%%%%%%%%%
In Ref.~\cite{PS}, we defined the $\pi \to \gamma$ leading twist 
transition distribution amplitudes (TDAs) from the 
matrix elements
\begin{equation}
   \langle \gamma|\, {\bar q}^{\alpha}(z_{1})\, 
[z_1;z_0]\,{ q}^{\beta}(z_{0}) \,|\pi \rangle \Big|_{z_{i}^+=0,\, 
{z}_{i}^T=0}
 \end{equation} 
where the Wilson line 
$[y;z]  \equiv   {\rm  P\   exp\,}  \left[ig(y-z)\int_0^1\!dt\,
\,n_\mu A^\mu  (ty+(1-t)z)\right]$ provides  the QCD-gauge
invariance for  non local operators  and equals unity in  a light-like
(axial) gauge. In a similar way, we shall define in section 2 the nucleon to meson TDAs 
from the matrix elements
\begin{equation}
   \langle \pi|\, {q}^{\alpha}(z_{1})\, 
[z_1;z_0]\,{q}^{\beta}(z_{2})\, [z_2;z_0]\,
   {q}^{\gamma}(z_{3})\,[z_3;z_0] \,|p \rangle \Big|_{z_{i}^+=0,\, 
{z}_{i}^T=0} .
 \end{equation}

The $\bar p N \to \gamma^* \pi $ 
amplitude at small momentum transfer is then proportional to the 
TDAs $T(x_{i}, \xi, t)$, 
where $x_i$ (i=1,2,3) denote the light cone
momentum fractions carried by participant quarks, and $\xi$ is the
skewedness parameter connected  with $x_B$ by
\begin{equation}
  \label{xi-vs-xB}
\xi \approx \frac{x_B}{2-x_B}
\end{equation}
in the Bjorken limit. It reads schematically
\begin{equation}
{\cal M} (Q^2,  \xi, t)= \int dx dy \phi(y_i,Q^2)
T_{H}(x_i, y_{i}, Q^2) T(x_{i}, \xi, t, Q^2)\;,
\label{amp}
\end{equation}
where $\phi(y_i,Q^2)$ is the antiproton distribution amplitude and
$T_{H}$ the
hard scattering amplitude, calculated in the colinear approximation.
We shall show in section 3 that these TDAs obey QCD evolution equations,
which, as always, follow from the renormalization group equation of 
an appropriate operator, in our case of the
three quark operator. 
Their $Q^2$
dependence is thus completely  under control.
  We calculate in
section 4 the hard amplitude and derive some phenomenological
model-independent predictions of our picture.
In section 5, we comment on  processes related by crossing, such as
$\pi N \to N' \gamma ^*$
and $\gamma ^* N \to N' \pi$ which may be experimentally studied at
intense meson beams
  facilities and at JLab or Hermes, respectively.

 \section{The  $N  \to   \pi$  transition  distribution
amplitude}

\noindent Let  us take  a closer look  at the  transition distribution
amplitudes  from a nucleon to a pseudoscalar meson.  A similar description of the
antiproton to meson TDA may be straightforwardly deduced from our study. For
their definition we introduce light-cone coordinates $v^\pm = (v^0 \pm
v^3) /\sqrt{2}$ and  transverse components $v_T = (v^1,  v^2)$ for any
four-vector $v$.   The skewedness variable  $\xi = -\Delta^+  /2P^+$ with $\Delta = p'-p$ and
$P=(p+p')/2$  describes  the loss  of  plus-momentum  of the  incident
hadron in the proton $\to$ meson transition. We parametrize the quark momenta as 
shown on Fig. 1. The fractions of + 
momenta are labelled $x_{1}$, $x_{2}$ and $x_{3}$, 
and their supports are within $[-1+\xi, 1+\xi]$.  
Momentum conservation implies (we restrict to the case $\xi > 0$ ) :
\begin{equation}
\sum_{i}  x_{i} = 2 \xi \, .
\label{Sum}
\end{equation}
The fields with positive momentum fractions, $x_i\ge 0$, describe creation of 
quarks, whereas those with  negative momentum fractions, $x_i\le 0$, the 
absorption of antiquarks.
The spinorial and Lorentz decomposition of the matrix element follows the same line as 
in the case of the baryon distribution amplitude \cite{BFMS, BDKM}. Because of that let us first 
recall the definition of the proton DA at leading twist

\bea{bDA}
&&4\langle 0|\epsilon^{ijk}u^i_\alpha(z_1\,n)u^j_\beta(z_2\,n)d^k_\gamma(z_3\,n)|B(p,s)\rangle 
 \\
&&= f_N \left[V(\hat p\,C)_{\alpha \beta} (\gamma^5\,B)_\gamma 
+  A (\hat p\,\gamma^5\,C)_{\alpha \beta} B_\gamma
+ T(p^\nu i\sigma_{\mu \nu}\,C)_{\alpha \beta} 
(\gamma^\mu\,\gamma^5\,B)_\gamma 
\right] \;,
\nonumber
\eea
where $i,j,k$  are color indices and $n$ is the light cone +
 direction. The vector $n^\mu$ is a  light-like vectors ($n^2=0$) 
which together with $p^\mu$ defines the light-cone kinematics.

We then define the leading twist TDAs for the $p \to \pi^0$ transition as
:
\bea{TDA}
%\begin{eqnarray}
%\label{TDA}
 &&  4  \langle     \pi^{0}(p')|\, 
\epsilon^{ijk}u^{i}_{\alpha}(z_1\,n) u^{j}_{\beta}(z_2\,n)d^{k}_{\gamma}(z_3\,n)
\,|p(p,s) \rangle \Big|_{z^+=0,\,  z_T=0}  
\\ \nonumber
&&= -\frac{f_N}{2f_\pi}\left[ V^{0}_{1} (\hat P C)_{\alpha\beta}(B)_{\gamma}  +
A^{0}_{1} (\hat P\gamma^5 C)_{\alpha\beta}(\gamma^5 B)_{\gamma} -
3\,T^{0}_{1} ( P^\nu i\sigma_{\mu\nu} C)_{\alpha\beta}(\gamma^\mu B)_{\gamma} \right] 
\nonumber \\
&&+ V^{0}_{2} 
 (\hat P C)_{\alpha\beta}(\hat \Delta_{T} B)_{\gamma} +
A^{0}_{2}(\hat P \gamma^5 C)_{\alpha\beta}(\hat \Delta_{T}\gamma^5 B)_{\gamma}
+ T^{0}_{2} (\Delta_{T}^\mu P^\nu \sigma_{\mu\nu} C)_{\alpha\beta}(B)_{\gamma}
\nonumber \\
&&+  T^{0}_{3} ( P^\nu \sigma_{\mu\nu} C)_{\alpha\beta}(\sigma^{\mu\rho}
\Delta_{T}^\rho B)_{\gamma} + \frac{T_{4}^0}{M } (\Delta_{T}^\mu P^\nu 
\sigma_{\mu\nu} C)_{\alpha\beta}(\hat \Delta_{T}¥B)_{\gamma}\;, \nonumber
%\end{eqnarray}
\eea
where $\sigma^{\mu\nu}= i/2[\gamma^\mu, \gamma^\nu]$, $C$ is the charge 
conjugation matrix 
and $B$ the nucleon spinor. 
$\hat P = P^\mu \gamma_\mu$, the vector $\Delta =p'-p$ has 
 - in the massless limit - the  transverse components 
$$
\Delta_T^\mu = (g^{\mu\nu} - \frac{1}{Pn}(P^\mu n^\nu +P^\nu n^\mu))\Delta_\nu\;.
$$
$f_\pi$ is the pion decay constant ( $f_\pi = 93$ MeV) and $f_N$ is the
constant which determines the value of the nucleon wave function at the
origin, and which has been estimated through QCD sum rules to be of
order $5.3\cdot 10^{-3}$ GeV$^2$ \cite{COZ}.
The first three 
terms in (\ref{TDA}) are the only ones surviving the forward limit
$\Delta_T \to 0$.
The constants in front of these three  
terms  have been chosen
in reference to  the soft pion limit results (see below).
 With these conventions each function $V(z_{i}P\cdot n)$, $A(z_{i}P\cdot n)$,
$T(z_{i}P\cdot n)$ is 
dimensionless.
 Finally let us note that the number of leading twist TDAs in (\ref{TDA})
corresponds to eight 
independent helicity amplitudes related to the matrix element in
(\ref{TDA}) ($2^4/2=8$).

For the $n \to \pi^-$ TDA the analogous expression has the form

\bea{TDAminus}                                                                                
%\begin{eqnarray}
%\label{TDAminus}
  &&  4  \langle     \pi^{-}(p')|\,
\epsilon^{ijk}u^{i}_{\alpha}(z_1\,n)
u^{j}_{\beta}(z_2\,n)d^{k}_{\gamma}(z_3\,n)
\,|n(p,s) \rangle \Big|_{z^+=0,\,  z_T=0} =
\nonumber \\
&&  \frac{f_N}{\sqrt{2} f_\pi}   \left[ V^{-}_{1} (\hat P
C)_{\alpha\beta}(B)_{\gamma}  +
A^{-}_{1} (\hat P\gamma^5 C)_{\alpha\beta}(\gamma^5 B)_{\gamma} +
T^{-}_{1} ( P^\nu \sigma_{\mu\nu} C)_{\alpha\beta}(\gamma^\mu
B)_{\gamma} \right]
\nonumber \\
&&+ V^{-}_{2}
  (\hat P C)_{\alpha\beta}(\hat \Delta_{T} B)_{\gamma} +
A^{-}_{2}(\hat P \gamma^5 C)_{\alpha\beta}(\hat \Delta_{T}\gamma^5
B)_{\gamma}
+ T^{-}_{2} (\Delta_{T}^\mu P^\nu \sigma_{\mu\nu}
C)_{\alpha\beta}(B)_{\gamma}
\nonumber \\
&&+  T^{-}_{3} ( P^\nu \sigma_{\mu\nu} C)_{\alpha\beta}(\sigma^{\mu\rho}
\Delta_{T}^\rho B)_{\gamma} + \frac{T_{4}^-}{M }
(\Delta_{T}^\mu P^\nu
\sigma_{\mu\nu} C)_{\alpha\beta}(\hat \Delta_{T} B)_{\gamma} \;. 
%\end{eqnarray}
\eea                                                                          
 One might
reexpress $V_i^0$ and  $V_i^-$ (respectively $A_i^0$ and  $A_i^-$,
respectively $T_i^0$ and  $T_i^-$) in terms of the isospin 1/2 and 3/2
quantities $V_i^{I=1/2}$ and  $V_i^ {I=3/2}$ (respectively $A_i^
{I=1/2} $ and  $A_i^ {I=3/2} $,  respectively $T_i^ {I=1/2} $ and
$T_i^ {I=3/2} $).
Simple isospin rotation enables to deduce the TDAs for $p \to \pi^0 $
and $n \to \pi^0 $ transitions from Eqs.~(\ref{TDA}) and (\ref{TDAminus}).
We do not write down the $p \to \pi^- $ transition (which is pure
isospin  3/2 exchange ) since it does not contribute to the process
under study.

Each TDA can then be Fourier transformed to get the usual representation in terms of the 
momentum fractions, through the relation
\begin{equation}
F (z_{i}P\cdot n) = \int\limits^{1+\xi}_{-1+\xi} d^3x 
\delta (x_{1}+ x_{2}+ x_{3} -2\xi) e^{-iPn\Sigma x_{i}z_{i}} \, F(x_{i},\xi)
\end{equation}
where $F$ stands for $V_{i}, A_{i}, T_{i}$ and $\int d^3x \equiv 
\int dx_1dx_2dx_3 \delta(2\xi - x_1 -x_2-x_3)$.

 \section{Evolution equations}
QCD radiative corrections lead as usual to logarithmic scaling violations. 
The scale dependence of the proton to meson  TDAs is governed by 
evolution equations which are an extension of the evolution equations for
usual DAs and GPDs \cite{ERBL, Dvcs, Dvcs2}. The  derivation of evolution equation for TDAs proceeds 
in an analogous way as for DAs and for GPDs therefore we sketch only essential steps.

The 
 non-local three-quark operators
relevant for TDAs and their evolution, see 
\cite{BDKM} for details and notation which we follow, 
 involve quark fields having definite chirality or helicity
\beq{arrows} 
q^{\uparrow (\downarrow)}= \frac{1}{2}\left(1\pm \gamma^5  \right)q\;.
\eeq
The separation of ``minus'' components of quark fields \cite{BFMS} 
 leading to the 
dominant  twist-2 contribution is 
 achieved by the substitution $q \to \hat n q$, with $\hat n=n^\mu \gamma_\mu$.
There are two relevant operators in our problem:
the first one  corresponds to the case where the three quarks have
total helicity $1/2$ 
\beq{relop1} 
B^{1/2}_{\alpha \beta \gamma}(z_1,z_2,z_3)= \epsilon^{ijk}(\hat n q_i^\uparrow)_\alpha(z_1 n)
(\hat n q_j^\downarrow)_\beta(z_2 n) (\hat n q_k^\uparrow)_\gamma(z_3 n)
\eeq
and  the second one with the total helicity  $3/2$
\beq{relop3} 
B^{3/2}_{\alpha \beta \gamma}(z_1,z_2,z_3)= \epsilon^{ijk}(\hat n q_i^\uparrow)_\alpha(z_1 n)
(\hat n q_j^\uparrow)_\beta(z_2 n) (\hat n q_k^\uparrow)_\gamma(z_3 n)\;.
\eeq
For simplicity, we have assumed that all quarks in (\ref{relop1}), (\ref{relop3}) have different flavours.
The conditions imposed by flavour symmetry do not influence the evolution equations but lead to 
certain  symmetry 
properties of TDAs.
Since operators (\ref{relop1}), (\ref{relop3}) belong to different representations of the 
Lorentz group they do not mix with each other.

The operators $B$, (\ref{relop1}) and (\ref{relop3}), satisfy the renormalisation group equation
\beq{rengr}
\mu \frac{d }{d\,\mu}\;B = H \cdot B \;
\eeq
with $H$ being an integral \cite{BDKM} 
operator\footnote{We restored in Eq.~(\ref{intop}) the factor $-\alpha_s/(2\pi)$ 
absent in Eq.~(2.23) of \cite{BDKM}.}
\beq{intop}
H = -\frac{\alpha_s}{2\pi}\left[ \left( 1+1/N_c \right) {\cal H} + 3C_F/2 \right]
\eeq
in which the second term $\sim C_F$ corresponds to the self-energy corrections of each quark field.
The operator ${\cal H}$ acts in different way on the operator (\ref{relop3}) and on the 
operator (\ref{relop1}). In the first case it is determined by
 contributions from  one loop Feynman diagrams 
describing  in the Feynman gauge the vertex corrections corresponding to gluon exchanges between 
quark fields and gluons forming Wilson lines as
\beq{H3}
{\cal H}_{3/2} = {\cal H}^v_{1\,2} + {\cal H}^v_{2\,3}  +  {\cal H}^v_{1\,3} \;,  
\eeq
where
\bea{Hv12}
&& {\cal H}^v_{1\,2} B(z_i) = - \int\limits_0^1 \frac{d\alpha}{\alpha} \left\{
\bar \alpha\left[B(z^\alpha_{1\,2},z_2,z_3)- B(z_1,z_2,z_3)    \right] \right.
\nonumber \\
&& \left. + \bar \alpha \left[B(z_1,z^\alpha_{2\,1},z_3)- B(z_1,z_2,z_3) \right] \right\}\;,
\eea
with $\bar \alpha = 1-\alpha$, $z^\alpha_{i\,k}=z_i\bar \alpha +z_k \alpha$.  
In the case of (\ref{relop1}) the operator ${\cal H}$ is determined not only by above 
contributions but also by those ones which correspond to Feynman diagrams with gluon 
exchange between quark lines having opposite chiralities (i.e. in our case between lines (1,2) 
and (2,3))
and it can be written as
\beq{H1}
{\cal H}_{1/2} = {\cal H}_{3/2} - {\cal H}^e_{1\,2} - {\cal H}^e_{2\,3}\;,   
\eeq
where
\beq{He12}
 {\cal H}^e_{1\,2} B(z_i) = \int {\cal D}\alpha \;B(z^{\alpha_1}_{1\,2},z^{\alpha_2}_{2\,1},z_3)\;,
\eeq
with 
\beq{measure}
\int\;{\cal D}\alpha = \int\limits_0^1\;d\alpha_1\,d\alpha_2\,d\alpha_3\,
\delta(1- \alpha_1 - \alpha_2 - \alpha_3)\;.
\eeq

From the renormalisation group equation (\ref{rengr}) and 
(\ref{intop})
we derive the corresponding equation for
the matrix element of operators $B$ between states relevant for the process under study. 
The subsequent use of 
parametrisations of these matrix elements in terms of different TDAs, 
as those given by Eqs.~(\ref{TDA}) or (\ref{TDAminus}), results in
 the evolution equations for TDAs $V_i$, $A_i$, $T_i$.

As a definite example we present the evolution equation for the set of TDAs denoted as 
$F^{\uparrow \downarrow \uparrow}(x_i)$ which are
related to the matrix element of three quark operator with total helicity $1/2$
\[ 
\langle \pi^0(p')| \epsilon^{ijk}(\hat n q_i^\uparrow)_\alpha(z_1 n)
(\hat n q_j^\downarrow)_\beta(z_2 n) (\hat n q_k^\uparrow)_\gamma(z_3 n)|N(p,s) \rangle
\]
being parametrised according to Eq.~(\ref{TDA}).
The evolution equation  has the form
\bea{evoleq}
%\begin{eqnarray}
%\label{evoleq}
&&Q\frac{d}{d\,Q}\;F^{\uparrow \downarrow \uparrow}(x_i) 
= -\frac{\alpha_s}{2\pi}\left\{
\frac{3}{2}\,C_F\,F^{\uparrow \downarrow \uparrow}(x_i) 
- \left(1+\frac{1}{N_c} \right)
\right. 
 \\
&&\left[\left(\int\limits_{-1+\xi}^{1+\xi}dx_1' 
\left[\frac{x_1\rho(x_1',x_1) }{x_1'(x_1' - 
x_1)}  \right]_+
+\int\limits_{-1+\xi}^{1+\xi}dx_2' 
\left[\frac{x_2\rho(x_2',x_2) }{x_2'(x_2' - 
x_2)}  \right]_+ \right)F^{\uparrow \downarrow \uparrow}(x_1',x_2',x_3)
\right.
\nonumber \\
&&+\left(\int\limits_{-1+\xi}^{1+\xi}dx_1' 
\left[\frac{x_1\rho(x_1',x_1) }{x_1'(x_1' - 
x_1)}  \right]_+
+\int\limits_{-1+\xi}^{1+\xi}dx_3' 
\left[\frac{x_3 \rho(x_3',x_3) }{x_3'(x_3' - 
x_3)}  \right]_+ \right) F^{\uparrow \downarrow \uparrow}(x_1',x_2,x_3')
\nonumber \\
&&+ \left(\int\limits_{-1+\xi}^{1+\xi}dx_2' 
\left[\frac{x_2\rho(x_2',x_2) }{x_2'(x_2' - 
x_2)}  \right]_+ 
+ \int\limits_{-1+\xi}^{1+\xi}dx_3' 
\left[\frac{x_3 \rho(x_3',x_3) }{x_3'(x_3' - 
x_3)}  \right]_+ \right)F^{\uparrow \downarrow \uparrow}(x_1,x_2',x_3')
\nonumber \\
&&+ \frac{1}{2\xi-x_3}\left( \int\limits_{-1+\xi}^{1+\xi}dx_1' 
\frac{x_1}{x_1'}\rho(x_1',x_1)    
+  \int\limits_{-1+\xi}^{1+\xi}dx_2' 
\frac{x_2}{x_2'}\rho(x_2',x_2)  
 \right)F^{\uparrow \downarrow \uparrow}(x_1',x_2',x_3)
\nonumber \\
&&\left. \left.
 +\frac{1}{2\xi - x_1} \left(
\int\limits_{-1+\xi}^{1+\xi}dx_2' 
\frac{x_2}{x_2'}\rho(x_2',x_2)  
+\int\limits_{-1+\xi}^{1+\xi}dx_3' 
\frac{x_3}{x_3'} \rho(x_3',x_3)   
\right) F^{\uparrow \downarrow \uparrow}(x_1,x_2',x_3')
\right] \right\}\;. \nonumber
%\end{eqnarray}
\eea
The integration region in each integral is restricted in two ways.
Firstly, the support of integrands is defined by functions
 $\rho(x,y)=\theta(x\ge y \ge 0)-\theta(x\le y \le 0)$, with
$\theta(x\ge y \ge 0) = \theta(x\ge y)\theta(y \ge 0)$. 
This function  $\rho(x,y)$ is a generalization of analogous one 
which appears in equations describing the pure evolution ERBL \cite{Dvcs}. 
The second condition is the requirement that although not
  denoted as a variable of integration 
the variables $x'_i$
must satisfy the condition that
$x'_i \in [-1+\xi,1+\xi]$, e.g. in the first integral 
over $x_1'$ on the rhs of (\ref{evoleq}) the variable $x_2'=2\xi - x_3 -x_1'$
must belong to the interval  $x_2' \in [-1+\xi,1+\xi]$.

The evolution equation for the set of TDAs  
$F^{\uparrow \uparrow \uparrow}(x_i)$, which correspond to the case where
 three quarks have total helicity $3/2$, is obtained from (\ref{evoleq}) by neglecting
two last lines.

The results of this evolution are different in the various $x_{i}$ sectors. 
In particular, when all  $x_{i} > 0$ 
 one is in the same kinematics as the usual ERBL equation for the 
baryons, with the simple $x_{i} \to x_{i}/2\xi$ rescaling. 
%The solutions of this equation are well known in terms of Appell polynomials
The solutions of the Eq.~(\ref{evoleq}) in this ERBL region are thus well known and are 
expressed in terms of  Appell polynomials
 $P_n(x_i /2\xi)$:
\beq{expansion}
%\begin{equation}
\hspace{-.3cm}F(x_i, \xi, \mu^2)=x_1 x_2 x_3 \delta (x_1+ x_2 +x_3 - 2\xi)\sum_n \phi_n\, P_n(x_i /2\xi)
\left (\frac{
\alpha_s(\mu)}{\alpha_s(\mu_0)}
\right )^{\gamma_n/b_0},
%\label{expansion}
%\end{equation}
\eeq
where $b_0=11/3 N_c-2/3n_f$,
$\gamma_n$ are the corresponding anomalous dimensions
and $\phi_n(\mu_0)$ are dimensionless nonperturbative parameters.

When one or two  $x_{i} < 0$ , the solutions of evolution equation (\ref{evoleq})   are unknown: 
they will deserve further study.
Although asymptotic solutions are simple, one should not use them
without caution in phenomenological studies, since it is known 
 that the asymptotic solution 
 of the proton distribution amplitude
 (namely,  the DA
proportional to $x_1x_2x_3 $ like the first term in (\ref{expansion})) 
does not allow for a good description of form factors at accessible
values of $Q^2$. Thus one should not insist on the
asymptotic form but instead
use nonperturbative techniques such as QCD sum rules or
lattice calculations to get boundary values to insert in the evolution
equations, and maybe solve them by applying some methods
 based on conformal symmetry used in \cite{BDKM} or by methods proposed in
\cite{MKS}.

 \section{Hard amplitude and cross section estimates}
 Since the leading twist antiproton distribution amplitude selects the helicity $\pm \frac {1} {2}$
 state for the hard scattered quarks, and since the photon coupling does not modify the helicity
 of these quarks, the three quarks extracted from the proton by the TDA have also a total helicity
 of $\pm \frac {1} {2}$. Moreover, in this first study, we concentrate on terms which are not 
 vanishing at zero $\Delta_{T}¥$, {\em i.e.} to the contributions of the three first TDAs in Eq. (\ref{TDA}), namely $V_{1}, A_{1},T_{1}$.
 The hard amplitude is then at leading twist straightforwardly deduced from
 the studies of proton form factors at high  $Q^2$. 
 At  leading order in $\alpha_{S}$, the amplitude  ${\cal M}^\mu $ for the reaction 
 $$
 \bar p(k, \lambda) p(p,s)\to \gamma^*(q) \pi^0(p') 
 $$
 may be read off from baryonic form factor calculations \cite{ERBL,CZ} as
\bea{hardamp}
%\begin{eqnarray}
{\cal M}^\mu &= & -i e_{p}¥F (Q^2,\xi,t)¥ \bar v(k,\lambda) \gamma^\mu \gamma^5 u(p,s)¥ 
\nonumber\\
F (Q^2,\xi,t) &=& \frac{f_{N}^2}{ f_{\pi}}¥ \frac{(4 \pi \xi \alpha_S(Q^2))^2}{27\,Q^4} \int\limits_{1+\xi}^{-1+\xi} d^3x \int\limits_0^1 d^3y 
\sum\limits_{\alpha=1}^{10} T_{\alpha}(x_{i},y_{j}¥)
%\label{hardamp}
%\end{eqnarray}
\eea
with
\bea{T}
%\begin{eqnarray}
T_{1}&=&\frac{4}{3}  \frac{{\cal V}+ {\cal T}}{(2\xi-x_{1}+i\epsilon)^2
(x_{3}+i\epsilon)(1-y_{1}¥)^2y_{3}}¥ \nonumber\\
T_{2}&=&-\frac{4}{3}  \frac{ {\cal T}}{(x_{1}+i\epsilon)
(2\xi-x_{2}+i\epsilon)(x_{3}+i\epsilon)y_{1}¥(1-y_{2}¥)y_{3}}¥ \nonumber\\
T_{3}&=&\frac{4}{3}  \frac{{\cal V}}{(x_{1}+i\epsilon)
(2\xi-x_{3}+i\epsilon)(x_{3}+i\epsilon)y_{1}¥(1-y_{1}¥)y_{3}}¥ \nonumber\\
T_{4}&=&-\frac{4}{3}  \frac{{\cal V}}{(x_{2}+i\epsilon)(2\xi-x_{3}+i\epsilon)
(x_{3}+i\epsilon)y_{2}¥(1-y_{1}¥)y_{3}}¥ \nonumber\\
T_{5}&=&-\frac{2}{3} \frac{{\cal V}}{(2\xi-x_{3}+i\epsilon)^2(1-y_3)^2}\left( 
\frac{1}{(x_{1}+i\epsilon)y_{1}} +
 \frac{1}{(x_{2}+i\epsilon)y_{2}} \right) 
\nonumber\\
T_{6}&=&\frac{2}{3}  \frac{{\cal V}+ {\cal T}}{(2\xi-x_{1}+i\epsilon)^2
(x_{2}+i\epsilon)(1-y_{1}¥)^2y_{2}}¥ \\
T_{7}&=&\frac{2}{3}  \frac{{\cal V}+ {\cal T}}{(2\xi-x_{1}+i\epsilon)^2
(x_{2}+i\epsilon)(1-y_{1}¥)^2y_{2}}¥ \nonumber\\
T_{8}&=&\frac{1}{3}  \frac{{\cal V}}{(x_{1}+i\epsilon)(x_{2}+i\epsilon)
(2\xi-x_{3}+i\epsilon)y_{1}¥(1-y_{1}¥)y_{2}}¥ \nonumber\\
T_{9}&=&-\frac{1}{3}  \frac{ {\cal T}}{(x_{1}+i\epsilon)
(2\xi-x_{1}+i\epsilon)(x_{2}+i\epsilon)y_{1}¥(1-y_{2}¥)y_{2}}¥ \nonumber\\
T_{10}&=&\frac{1}{3}  \frac{{\cal V}}{(x_{1}+i\epsilon)(x_{2}+i\epsilon)
(2\xi-x_{1}+i\epsilon)x_{3}y_{1}y_{2}(1-y_{3}¥)}¥ \nonumber
%\end{eqnarray}
\eea
with
 \begin{equation}
{\cal V}(x_{j},y_{i},\xi, t) = [V(y_{i}) - A(y_{i}) ] \cdot 
[V_{1}(x_{j},\xi, t) - A_{1}(x_{j},\xi, t) ]\;,
\end{equation}
\begin{equation}
{\cal T}(x_{j},y_{i},\xi, t) = -12 [T(y_{i})  ] . [T_{1}(x_{j},\xi, t) ]\;,
\end{equation}
and $\int d^3y \equiv \int dy_1dy_2dy_3\delta(1-y_1-y_2-y_3)$.
%\section{Cross section estimates}
Note that the  electromagnetic vector current has been replaced in the
amplitude  (\ref{hardamp})  by an axial vector current, 
due to the presence of the outgoing
pseudo scalar meson $\pi$-meson in the TDA (\ref{TDA}).

A model independent result of our analysis 
is the  scaling law for  the amplitude :
\begin{equation} 
{\cal M}(Q^2, \xi) \sim \frac{\alpha_s(Q^2)^2}{Q^4}\;,
\label{scaling}
\end{equation}
valid up to  logarithmic corrections due to DA and TDA anomalous dimensions. 
On the other hand the ratio :
$$
\frac {d\sigma(\bar p p \to l^+ l^- \pi^0) / dQ^2}  {d\sigma(\bar p p \to l^+
l^- ) / dQ^2} \;,
$$
should be almost $Q^2$ independent. 

Another interesting consequence of our framework is the dominance of the transverse 
polarization of the virtual photon, which results in a specific angular distribution of the
 lepton pair in its rest frame, namely 
  \begin{equation} 
\frac {d\sigma (p \bar p \to l^+ l^- \pi)}{\sigma d\theta} \sim 1+ cos^2 \theta
\label{ang}
\end{equation}
 where $\theta$ is the usual emission angle of the lepton in the virtual photon center 
 of mass frame.
  
Since the electromagnetic form factor $F_1^p$ is obtained with the same
expressions as Eqs.~(\ref{hardamp}), (\ref{T}) with
the replacement of the axial vector current by vector one and with
$2 f_\pi \to 1$, and ${\cal T} \to 4  [T(y_{i})  ] \cdot [T(x_{j}) ]$
one may obtain an estimate of the threshold cross section in terms of
the electromagnetic form factor , once a reasonable form of $V$, $A$ and $T$
are chosen. We shall not enter more the phenomenology in this letter
but will study it in a forthcoming work.

\vspace{0.3cm}
\section{Related processes and conclusions}

\noindent 
We have defined the new transition distribution amplitudes 
$N \to \pi $, i.e. which parametrize the matrix elements of 
light-cone operators between  a baryon and a meson states; 
this  generalizes the concept of GPDs for non-diagonal transitions. 
Similar matrix element was introduced also in Ref.~\cite{FPPS}
in discussion of the exclusive production of
forward baryons off nucleons.
Obviously related processes 
are 

\begin{itemize}
\item Firstly, the exclusive lepton pair production
\begin{equation} 
\pi N \to N' \gamma ^*\;,
\label{crossed1}
\end{equation} 
in the kinematical regime where the outgoing nucleon is almost colinear 
to the incoming 
meson, which is the backward region of the reaction $\pi N \to  \gamma ^* N'$
 studied in Ref.\cite{BDP}, 
and which may be studied in an intense pion beam facility such
  as the project JPark. 
  
\item
  Secondly, the same framework may be applied to
 \begin{equation} 
\gamma ^* N \to N' \pi\;,
\label{crossed2}
\end{equation} 
in the kinematical regime where the outgoing meson is almost colinear to the incoming 
proton, which is the spacelike analog of the previous case. 

\item Thirdly, the crossed ($t\to s $) version of the $N \to \pi$ TDA 
describes the exclusive 
fragmentation of three colinear quarks in a Baryon-meson pair.
 Analogously to the two
meson generalized distribution amplitude \cite{Dvcs}, it is an 
interesting tool 
to access reactions where an emerging isolated baryon is replaced by 
a baryon-meson system.
An exemple of its usefulness is the calculation of the impact factor of 
a hard diffractive 
reaction 
  (at large values of $t$) 
where a baryon projectile is transformed into a baryon-meson ejectile. In this way, the 
crossed TDA describes the partonic content of a continuum or resonating baryon-meson state.

\end{itemize}

Many other processes may be studied in the same way, where the $\pi$ meson is replaced 
with  other mesons, or/and the outgoing nucleon is replaced with other baryons.

%\section{Conclusions}

The introduction of transition distribution amplitudes thus allows to
study the reactions
$$
 \bar p  p \to l^+ l^- \pi^0  ~~~~~~~~~  \bar p n \to l^+ l^- \pi^-
$$
along the same lines as the timelike electromagnetic baryon form factors.
 The applicability of a perturbative QCD approach to
the proton form factor at accessible energies has been controversial for
years \cite{contr}. We pretend here neither that experimental data have
shown that the perturbative approach is successful,  nor that next to
leading order corrections will succeed in describing them. We just want to
point out that a new phenomenology of related but different reactions is
now possible to try to understand the present puzzle. Moreover, the study
of form factors  has been the subject  of  
 many developments after the pioneering papers \cite{ERBL,CZ}. In
particular, the calculation of next to leading logarithm
corrections \cite{NLL}, the proposal of an optimal renormalization scale
fixing \cite{BLM},  the importance of soft gluon resummation \cite{SLi} and
the discussion of the timelike vs spacelike aspects \cite{GP}, have put to
a firmer basis the QCD description of these objects(for a short review,
see \cite{dJP}). These interesting developments ought to be applied now to
the meson production processes. This necessary effort should help a
phenomenological analysis to be more constrained so that one may clearly
see if a (quite sophisticated) perturbative QCD approach is indeed
relevant to the kinematical range which may be accessible at GSI-FAIR, with
$\sqrt{s}\approx 32$GeV$^2$ in the target rest frame 
and $\sqrt{s}\approx 207$GeV$^2$ in the accelerator mode. 
 On
the other hand, an alternative more phenomenological point of view has
been also developped to describe form factors in terms of generalized
parton distributions and the so-called "Feynman mechanism" \cite{FM}. 
This
framework has recently been enlarged \cite{KS}
 to describe reactions (\ref{pbarp}) at fixed angle through generalized
distribution amplitudes \cite{GDA}.

\vspace{0.3cm}
\noindent {\bf Acknowledgments.}

\noindent  We acknowledge useful  discussions and  correspondence 
with I.V.~Anikin, V.M.~Braun,
    J.C.~Collins, M.~Diehl, 
M.~D\"uren, G.P.~Korchemsky, T.N.~Pham, P.V.~Po\-by\-li\-tsa, 
M.V.~Polyakov, C.~Roiesnel and S.~Wallon.   
This  work  is  supported  by  
the Polish Grant 1 P03B 028 28, 
the
French-Polish scientific agreement Polonium and the Joint Research
Activity "Generalised Parton Distributions" of the european I3 program
Hadronic Physics, contract RII3-CT-2004-506078. 
L.Sz. is a Visiting Fellow of 
the Fonds National pour la Recherche Scientifique (Belgium).

%%%%%%%%%%%%%%%%%%%%%%%%%%%%%%%%%%%%%%%%%%%%%%%%%%%%%%%%%%%%%%%%%%%%%%%%%%%%

\end{document}